\documentstyle[epsfig]{aipproc}

\begin{document}
\title{Reconstructing Supersymmetric Theories at the Linear Collider\\ }

\author{Grahame A. Blair}
\address{Royal Holloway, Univ. of London, Egham, Surrey TW20 0EX,UK}

\maketitle

\begin{abstract}
High precision measurements at the linear collider will allow
a model-independent reconstruction of nature at high energy scales. 
The method of bottom-up extrapolation from the electroweak scale
to the GUT scale is explained and both a universal minimal supergravity
and a gaugino-mediated model are presented as examples. 
Comparisons are made with the LHC-only case.
\end{abstract}

\subsection*{Introduction} 
If supersymmetry (SUSY) is realised in nature, one of the major goals of the 
linear collider (LC) will be to determine how this symmetry is broken
\cite{ghodbole_zerwas}.  This problem is equivalent to determining the
structure of the theory at high energy scales where the
mechanism of SUSY breaking influences directly the parameters of the theory.
This report presents a model-independent approach where the SUSY parameters 
are reconstructed via extrapolations from the electroweak scale to the
GUT scale, assuming the precision measurements that may be performed
at a high luminosity e$^+$e$^-$ LC \cite{ee_params}.  More details
are presented in Ref.~\cite{our_paper}.

\subsection*{The Bottom-Up Approach}

A widely employed method to determine the fundamental SUSY parameters 
at the LHC~\cite{lhc} and the LC\cite{martyn} is
to assume a SUSY breaking scenario and then fit to the corresponding
experimentally determined low-energy particle spectrum.  While this
approach gives a useful indication of the SUSY measurement potential, 
the scenario assumptions are effectively constraints in the fit
and so may give a false impression.   This danger
is particularly present for models with pseudo-fixed point
structures, where the low-energy effective theories will be quite
similar for a range of fundamental parameters.  Additionally,
nature may not be regular at the GUT scale or may possess new intermediate
scales that are not immediately apparent from a top-down approach.

For these reasons a model-independent method is adopted
where the structure of the theory is extrapolated via the renormalisation
group equations (RGEs) from  low-energy to high energy, 
with input to the RGEs from experimental measurement alone.  In this 
bottom-up approach new intermediate scales may indeed become apparent, in 
which case the RGEs would need to adjusted accordingly.  An example
of such a case is the gauge mediated scenario that was addressed in
Ref.~\cite{our_paper}.  The bottom-up approach manifests the
quality of the reconstruction in a transparent form and stresses
the need for high accuracy measurements, especially in those
cases where universality at the GUT scale may be only slightly broken.

\subsection*{The Models and Experimental Input}

Presented here are two studies. Model A is 
minimal supergravity\cite{nilles} with 
parameters M$_\frac{1}{2}$=190 GeV, M$_0$=200 GeV,
A$_0$=550 GeV, $\tan\beta$=30, sign($\mu$)=$-$.  Model B is inspired by
the gaugino-mediated scenario~\cite{gaugino} with 
M$_\frac{1}{2}$=200 GeV, M$_0$=5 GeV,
A$_0$=0 GeV, $\tan\beta$=2.5, M$_{H_1}$=300 GeV, M$_{H_2}$=200 GeV,
sign($\mu$)=$-$.  A gauge-mediated model is presented in 
Ref.~\cite{our_paper}.

The experimental input to this study consists of particle masses
and polarized production cross-sections.  The analysis
requires a total integrated luminosity of about 1 ab$^{-1}$.  
For the particle mass
precisions, threshold scans are assumed such as discussed 
in Ref.~\cite{martyn}.  For the squarks and gluino, a generic precision
of 10 GeV is assumed from the LHC~\cite{lhc}.
Only statistical errors are included 
for the cross-sections and the polarisations are taken as
80\% for e$^-$ and 60\% for e$^+$.  
For model A, where the large $\tan\beta$
gives rise to multi-tau final states, we assume a reconstruction
efficiency of 20\% and inflate the errors on masses and cross-sections
accordingly.  For model B we assume a reconstruction efficiency of 80\%
and also allow for the fact that the sneutrinos happen to decay invisibly.

\begin{figure}[b!] 
\centerline{\epsfig{file=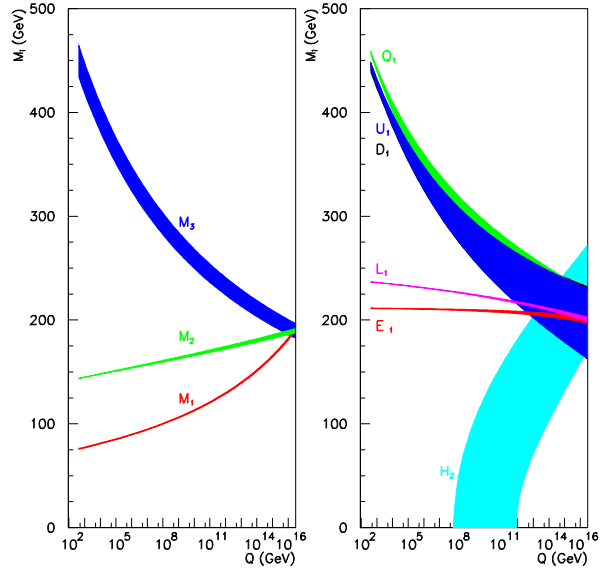,height=3.4in,width=3.4in}}
\vspace{10pt}
\caption{Extrapolation of model A gaugino (left) and
scalar (right) soft breaking parameters from the electroweak
scale to the GUT scale assuming a combination of LC and LHC errors.}
\label{lc_and_lhc_sugra}
\end{figure}

\begin{figure}[b!] 
\centerline{\epsfig{file=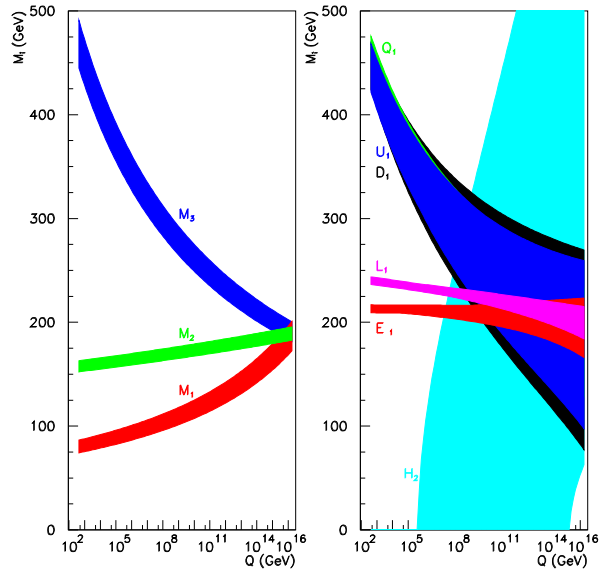,height=3.4in,width=3.4in}}
\vspace{10pt}
\caption{Model A, assuming generic LHC-only errors.  The uniformity
at the GUT scale is not so apparent.}
\label{lhc_sugra}
\end{figure}

The results of the extrapolations of the gaugino and scalar soft
breaking terms for model A, using LC and LHC data, are shown
in Fig.~\ref{lc_and_lhc_sugra}.  It is immediately clear that the
high precision slepton measurements at the LC give excellent evidence
for uniformity at the GUT scale.  The corresponding results
assuming generic LHC-only mass errors of 3 GeV for the non-coloured states 
are shown in Fig.~\ref{lhc_sugra}.  The effect of losing the 
complementary LC precision data is clear.

\begin{figure}[h] 
\centerline{\epsfig{file=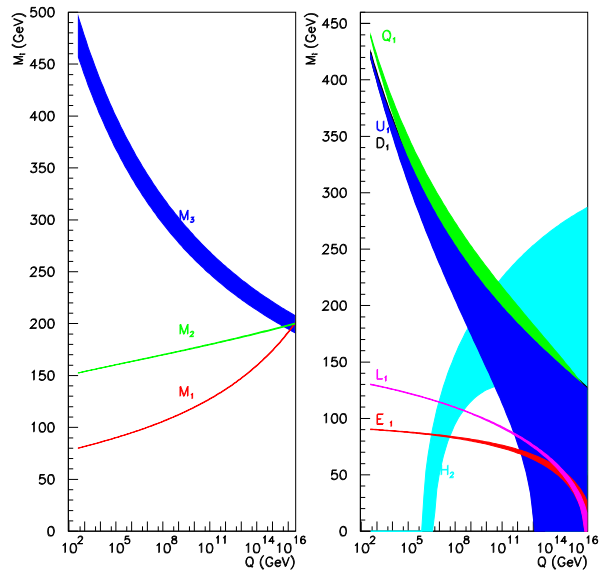,height=3.4in,width=3.4in}}
\vspace{10pt}
\caption{Extrapolation to the GUT scale for model B, assuming
LHC and LC data.  The result is clearly distinguishable from model A.}
\label{gaugino_mediated}
\end{figure}

The corresponding LC+LHC extrapolations for model B are shown in
Fig.~\ref{gaugino_mediated} where the no-scale structure is
apparent at the GUT-scale.  Clearly this approach can distinguish
between various scenarios, without any {\it a priori} model-dependent
assumptions.

\subsection*{Conclusion}

The bottom-up approach is a model-independent method of extrapolating
low-energy measurements to higher scales.  It avoids any
assumptions about high-energy structure and instead uses the experimental 
input alone to reconstruct the theory.
Complementing the LHC data with high precision measurements from the LC
will provide an excellent extrapolated view of physics at the GUT scale.

I would like to thank my collaborators, U.~Martyn, W.~Porod and P.~M.~Zerwas
and gratefully to acknowledge support from the British Council 
and from DESY, where this work was carried out.

\end{document}